\documentclass[runningheads]{llncs}
\usepackage{newclude}
\usepackage{graphicx}
\usepackage{amsfonts}
\usepackage{enumerate}
\usepackage{units}
\usepackage{comment}
\usepackage{color}
\usepackage{textcomp}
\usepackage{hyperref}
\usepackage{multirow}
\usepackage{environ}
\usepackage[boxed]{algorithm2e}
\usepackage{environ}
\usepackage{comment}
\usepackage{makecell}
\usepackage[classfont=roman,funcfont=roman,langfont=roman]{complexity}
\usepackage{tikz}
\usepackage{todonotes}
\usepackage[thmmarks]{ntheorem}
\usepackage{color}
\usepackage{transparent}
\usepackage{textgreek}

\newtheorem{observation}{Observation}
\theoremseparator{.}
\renewtheorem{claim}{Claim}
\renewtheorem{theorem}{Theorem}
\renewtheorem{lemma}{Lemma}
\theorembodyfont{\normalfont}
\renewtheorem{definition}{Definition}
\renewtheorem{corollary}{Corollary}
\theoremstyle{nonumberplain}
\theoremheaderfont{\itshape}
\theorembodyfont{\normalfont}
\theoremseparator{.}
\theoremsymbol{}
\theoremsymbol{\ensuremath{\square}}
\newtheorem{claimproof}{Proof of the claim}
\renewtheorem{proof}{Proof}

\newcommand{\MHVfull}{\textsc{Maximum Happy Vertices}}
\newcommand{\MHE}{\textsc{MHE}}
\newcommand{\MHV}{\textsc{MHV}}
\newcommand{\MHEfull}{\textsc{Maximum Happy Edges}}

\usepackage{tabularx,lipsum,environ,amsmath,amssymb}
\usepackage{mathtools}

\makeatletter
\newcommand{\problemtitle}[1]{\gdef\@problemtitle{#1}}
\newcommand{\probleminput}[1]{\gdef\@probleminput{#1}}
\newcommand{\problemquestion}[1]{\gdef\@problemquestion{#1}}
\NewEnviron{problemx}{
	\problemtitle{}\probleminput{}\problemquestion{}
	\BODY
	\par\addvspace{.5\baselineskip}
	\noindent
	\centering
	\boxed
	{
		\begin{tabularx}{330pt}{@{\hspace{0.5cm}} l p{250pt} c}
			\multicolumn{2}{@{\hspace{0.5cm}}l}{\@problemtitle} \\
			\textbf{Input:} & \@probleminput \\
			\textbf{Question:} & \@problemquestion
	\end{tabularx}}
	\par\addvspace{.5\baselineskip}
}

\renewcommand{\O}{\mathcal{O}}
\newcommand{\Ostar}[1]{\O^*(#1)}
\DeclareMathOperator*{\argmax}{argmax}
\DeclareMathOperator{\cw}{cw}

\newif\ifshort
\ifshort
	\makeatletter
	\newenvironment{lemmaO}[1][]{
		\if\relax\detokenize{#1}\relax
		\expandafter\@firstoftwo
		\else
		\expandafter\@secondoftwo
		\fi
		{\begin{lemma}[$\star$]}{\begin{lemma}[#1, $\star$]}
	}
	{
		\end{lemma}
	}
	\makeatother
	\makeatletter
\newenvironment{theoremO}[1][]{
	\if\relax\detokenize{#1}\relax
	\expandafter\@firstoftwo
	\else
	\expandafter\@secondoftwo
	\fi
	{\begin{theorem}[$\star$]}{\begin{theorem}[#1, $\star$]}
		}
		{
		\end{theorem}
	}
	\makeatother
\makeatletter
\newenvironment{claimO}[1][]{
	\if\relax\detokenize{#1}\relax
	\expandafter\@firstoftwo
	\else
	\expandafter\@secondoftwo
	\fi
	{\begin{claim}[$\star$]}{\begin{claim}[#1, $\star$]}
		}
		{
		\end{claim}
	}
	\makeatother
	\makeatletter
\newenvironment{corollaryO}[1][]{
	\if\relax\detokenize{#1}\relax
	\expandafter\@firstoftwo
	\else
	\expandafter\@secondoftwo
	\fi
	{\begin{corollary}[$\star$]}{\begin{corollary}[#1, $\star$]}
		}
		{
		\end{corollary}
	}
	\makeatother
	\NewEnviron{proofO}{}
	\NewEnviron{claimproofO}{}
\else

	\makeatletter
	\newenvironment{proofO}[1][]{\begin{proof}}{\end{proof}}
	\newenvironment{claimproofO}[1][]{\begin{claimproof}}{\end{claimproof}}
	\makeatother
	
\fi

\begin{document}
\title{Maximizing Happiness in Graphs of Bounded Clique-Width\thanks{This research was supported by the Russian Science Foundation (project 16-11-10123-\textPi)}}
%
%
\author{Ivan Bliznets\orcidID{0000-0003-2291-2556} \and
	Danil Sagunov\orcidID{0000-0003-3327-9768}}
\authorrunning{I. Bliznets and D. Sagunov}
%
\institute{St. Petersburg Department of Steklov Mathematical Institute of Russian Academy of Sciences, 27 Fontanka, St.\ Petersburg, Russia\\
	\email{\{iabliznets,danilka.pro\}@gmail.com}}

	\maketitle

\begin{abstract}

Clique-width is one of the most important parameters that describes structural complexity of a graph.
Probably, only treewidth is more studied graph width parameter. 
In this paper we study how clique-width influences the complexity of the \MHVfull~(\MHV)~and \MHEfull~(\MHE)~problems.
We answer a question of Choudhari and Reddy '18 about parameterization by the distance to threshold graphs by showing that \MHE~is \NP-complete on threshold graphs.
Hence, it is not even in \XP~when parameterized by clique-width, since threshold graphs have clique-width at most two.
As a complement for this result we provide a $n^{\O(\ell \cdot \cw)}$ algorithm for \MHE, where $\ell$ is the number of colors and $\cw$ is the clique-width of the input graph.
We also construct an \FPT~algorithm for \MHV~with running time $\Ostar{(\ell+1)^{\O(\cw)}}$, where $\ell$ is the number of colors in the input.
Additionally, we show $\O(\ell n^2)$ algorithm for \MHV~on interval graphs.

\end{abstract}

\section{Introduction}

Clique-width is one of the most important parameters that describe structural complexity of a graph.
Probably, only treewidth is more studied graph width parameter.
We note that one can treat clique-width as some generalization of treewidth as graphs of bounded treewidth have bounded clique-width.
Hence, the existence of an \FPT~algorithm parameterized by clique-width is a stronger result than the existence of an \FPT~algorithm parameterized by treewidth. 
Complexity of many problems were studied parameterized by the clique-width parameter, including \textsc{Max-Cut}~\cite{fomin2014almost},  \textsc{Edge Dominating Set}~\cite{fomin2014almost}, \textsc{Hamiltonian Path}~\cite{espelage2001solve}, \textsc{Graph $k$-Colorability}~\cite{gerber2003algorithms, kobler2001polynomial},  computation of the Tutte polynomial~\cite{gimenez2005computing}, \textsc{Dominating Set}~\cite{kobler2001polynomial, kobler2003edge}, computation of chromatic polynomial~\cite{makowsky2006computing}, and \textsc{Target Set Selection}~\cite{hartmann2018target}.
In this paper, we continue the line of the research and investigate computational and parameterized complexity of the \MHVfull~and \MHEfull~problems parameterized by clique-width of the input graph.

Before defining \MHVfull~and \MHEfull, we need to define what a happy vertex or a happy edge is.
\begin{definition}
	Let $G$ be a graph and let $c:V(G)\to[\ell]$ be a coloring of its vertices.
	We say that an edge $uv \in E(G)$ \emph{is happy with respect to $c$} (or simply \textit{happy}, if $c$ is clear from the context) if its endpoints share the same color, i.e.\ $c(u)=c(v)$.
	We say that a vertex $v \in V(G)$ \emph{is happy with respect to $c$} if all its neighbours have the same color as $v$, i.e.\ $c(v)=c(u)$ for each neighbour $u$ of $v$ in $G$.
\end{definition}
We now give the formal definition of both problems.

\begin{problemx}
	\problemtitle{\MHVfull~(\MHV)}
	\probleminput{A graph $G$, a partial coloring of vertices $p: S \rightarrow [\ell]$ for some $S\subseteq V(G)$ and an integer $k$.}
	\problemquestion{Is there a coloring $c: V(G) \rightarrow [\ell]$ extending partial coloring $p$ such that the number of happy vertices with respect to $c$ is at least $k$?}
\end{problemx}

\begin{problemx}
	\problemtitle{\MHEfull~(\MHE)}
	\probleminput{A graph $G$, a partial coloring of vertices $p: S \rightarrow [\ell]$ for some $S\subseteq V(G)$ and an integer $k$.}
	\problemquestion{Is there a coloring $c: V(G) \rightarrow [\ell]$ extending partial coloring $p$ such that the number of happy edges with respect to $c$ is at least $k$?}
\end{problemx}

\textsc{Maximum Happy Vertices} and \textsc{Maximum Happy Edges}  were introduced by Zhang and Li in 2015~\cite{zhang2015algorithmic}, motivated by their study of algorithmic aspects of homophyly law in large networks.
These problems recently attracted a lot of attention from different lines of reseach.
From the parameterized point of view, the problems were studied in~\cite{Agrawal2018,Aravind2016,Aravind2017,Choudhari2018,Misra2018,bliznets2019lower,bliznets2019on}.
Works \cite{zhang2015algorithmic,zhang2018improved, zhang2015improved,xu2016submodular} are devoted to approximation algorithms for \MHV~and \MHE.
Finally, Lewis et al.~\cite{lewis2019finding} study the problems from experimental perspective.

Before we state our results we mention some previously known results under different parameterizations.
Aravind et al.~\cite{Aravind2017} constructed  $\Ostar{\ell^{\operatorname{tw}}}$ and $\Ostar{2^{\operatorname{nd}}}$ algorithms for both \MHV~and \MHE, where $\operatorname{tw}$ is the treewidth of the input graph and $\operatorname{nd}$ is the neighbourhood diversity of the input graph.
Misra and Reddy \cite{Misra2018} constructed $\Ostar{\operatorname{vc}^{\O(\operatorname{vc})}}$ algorithms for both problems, where $\operatorname{vc}$ is the vertex cover number of the input graph.

\textbf{Our results:} Below $\operatorname{cw}$ is the clique-width of the input graph, $\ell$ is the number of colors in the input precoloring and $n$ is the number of vertices in the input graph.
In the paper we prove the following results for the \MHEfull~problem:
\vspace{-5pt}
\begin{itemize}
\item \MHE~admits an \XP-algorithm with running time $n^{\O(\ell \cdot \operatorname{cw})}$, if a $\cw$-expression of the input graph is given;
\item \MHE~does not admit an \XP-algorithm parameterized by clique-width alone, unless $\P=\NP$ (by showing that \MHE~is \NP-complete on threshold graphs).
\end{itemize}

\vspace{-3pt}
Note that the question of the complexity of \MHE~on the class of threshold graphs was asked explicitly by Choudhari and Reddy in~\cite{Choudhari2018}.

For the \textsc{Maximum Happy Vertices} problem we establish the following results:
\vspace{-3pt}
\begin{itemize}
\item \MHV~admits an \FPT~algorithm with $\Ostar{(\ell+1)^{\O(\cw)}}$ running time, if a $\cw$-expression of the input graph is given (note that \MHV~parameterized by clique-width alone is \W[2]-hard \cite{bliznets2019on});
\item Additionally, \MHV~is solvable on the class of interval graphs in time $\O(\ell n^2)$.
\end{itemize}


Our work shows that clique-width is a parameter under which computational complexity of problems \MHV~and \MHE~differ most significantly.
On graphs of bounded clique-width, \MHV~admits an FPT algorithm with running time $\Ostar{(\ell+1)^{\O(\cw)}}$, while \MHE~is \NP-complete on graphs of clique-width two and does not admit even an \XP-algorithm when parameterized by $\text{cw}$, however, we show that there is an \XP-algorithm for the extended parameter $\text{cw}+\ell$.
Note that when parameterized by treewidth, neighbourhood diversity or vertex cover, the problems are known to have similar complexity.
We believe that the \FPT~algorithm for \MHV~parameterized by $\text{cw}+\ell$ is the most interesting result of this paper.

After establishing existence of polynomial algorithms for problems on graphs of bounded clique-width, it is natural to investigate complexity of problems on minimal hereditary classes of unbounded clique-width.
Unit interval graphs is one of such graph classes~\cite{lozin2007clique}.
We show that \MHV~is polynomially solvable on the class of interval graphs, which is a wider graph class.
So we think that our result for interval graphs nicely complements our understanding of computational complexity of \MHV~parameterized by clique-width.
We note that interval graphs also separate \MHV~and \MHE, as \MHE~is \NP-complete on threshold graphs, that are a subclass of interval graphs.

	\section{Preliminaries}

\textbf{Basic notation.} We denote the set of positive integer numbers by $\mathbb{N}$.
For each positive integer $k$, by $[k]$ we denote the set of all positive integers not exceeding $k$, $\{1,2,\ldots, k\}$.
We use $\sqcup$ for the disjoint union operator, i.e.\ $A\sqcup B$ equals $A\cup B$, with an additional constraint that $A$ and $B$ are disjoint.

We use the traditional $\O$-notation for asymptotical upper bounds.
We additionally use the $\mathcal{O}^*$-notation that hides polynomial factors.
We investigate \MHV~and \MHE~mostly from the parameterized point of view.
For a detailed survey in parameterized algorithms we refer to the book of Cygan et al.\ \cite{cygan2015parameterized}.
Throughout the paper, we use standard graph notation and terminology, following the book of Diestel \cite{diestel2018graph}.
All graphs in our work are undirected simple graphs.

\textbf{Graph colorings.} When dealing with instances of \MHV~or \MHE, we use a notion of colorings.
A \textit{coloring} of a graph $G$ is a function that maps vertices of the graph to the set of colors.
If this function is partial, we call such coloring \emph{partial}.
If not stated otherwise, we use $\ell$ for the number of distinct colors, and assume that colors are integers in $[\ell]$.
A partial coloring $p$ is always given as a part of the input for both problems, along with graph $G$.
We also call $p$ a \emph{precoloring} of the graph $G$, and use $(G,p)$ to denote the graph along with the precoloring.
The goal of both problems is to extend this partial coloring to a specific coloring $c$ that maps each vertex to a color.
We call $c$ a \emph{full coloring} (or simply, a coloring) of $G$ that extends $p$.
We may also say that $c$ is a coloring of $(G,p)$.
For a full coloring $c$ of a graph $G$ by $\mathcal{H}(G,c)$ we denote the set of all vertices in $G$ that are happy with respect to $c$.

\textbf{Clique-width.}
In order to define \emph{cliquewidth}  we follow definitions presented by Lackner et al.\ in their work on \textsc{Multicut} parameterized by clique-width \cite{Lackner2012}.

To define clique-width, we need to define $k$-expressions first.
For any $k \in \mathbb{N}$, a \emph{$k$-expression} $\Phi$ describes a graph $G_\Phi$, whose vertices are labeled with integers in $[k]$.
$k$-expressions and its corresponding graphs are defined recursively.
Depending on its topmost operator, a $k$-expression $\Phi$ can be of four following types.

\begin{enumerate}
	\item \emph{Introducing a vertex.}
	$\Phi=i(v)$, where $i \in [k]$ is a label and $v$ is a vertex.
	$G_\Phi$ is a graph consisting of a single vertex $v$ with label $i$, i.e.\ $V(G_\Phi)=\{v\}$.
	\item \emph{Disjoint union.}
	$\Phi=\Phi'\oplus\Phi''$, where $\Phi'$ and $\Phi''$ are smaller subexpressions.
	$G_\Phi$ is a disjoint union of the graphs $G_{\Phi'}$ and $G_{\Phi''}$, i.e.\ $V(G_\Phi)=V(G_{\Phi'})\sqcup V(G_{\Phi''})$ and $E(G_\Phi)=E(G_{\Phi'})\sqcup E(G_{\Phi''})$.
	The labels of the vertices remain the same.
	\item \emph{Renaming labels.}
	$\Phi=\rho_{i\to j}(\Phi')$.
	The structure of $G_\Phi$ remains the same as the structure of $G_{\Phi'}$, but each vertex with label $i$ receives label $j$.
	\item \emph{Introducing edges.}
	$\Phi=\eta_{i,j}(\Phi')$.
	$G_\Phi$ is obtained from $G_{\Phi'}$ by connecting each vertex with label $i$ with each vertex with label $j$.	
\end{enumerate}

\emph{Clique-width} of a graph $G$ is defined as the smallest value of $k$ needed to describe $G$ with a $k$-expression and is denoted as $\operatorname{cw}(G)$, or simply $\cw$.

There is still no known \FPT-algorithm for finding a $k$-expression of a given graph $G$.
However, there is an \FPT-algorithm that decides whether $\operatorname{cw}(G) > k$ or outputs $(2^{3k+2}-1)$-expression of $G$.
For more details on clique-width we refer to \cite{Hlineny2007}. 

	


	\section{Maximum Happy Edges}

This section is dedicated to the \MHEfull~problem parameterized by clique-width.
We start with showing that \MHEfull~is \NP-complete on graphs of clique-width at most two.

In \cite{Choudhari2018}, Choudhari and Reddy proved that \MHV~is polynomially solvable on the class of threshold graphs (that have clique-width at most two \cite{zbMATH00803309}) and questioned the complexity of \MHE~on the same graph class.
We answer their question by showing that \MHEfull~is \NP-complete on threshold graphs.
To prove this, we require the following useful characterization of threshold graphs.

\begin{lemma}[\cite{Mahadev9780444892874}]\label{lemma:threshold_char}
	Threshold graphs are graphs that can be partitioned in a clique $K=\{u_1, u_2, \ldots, u_k\}$ and an independent set $I$, such that $N[u_i] \subseteq N[u_{i+1}]$ holds for every $i \in [k-1]$.
\end{lemma}

We now prove the abovementioned hardness of \MHE.

\begin{theorem}\label{thm:mhe_np_hard}
	\MHEfull~is \NP-complete on the class of threshold graphs.
\end{theorem}
\begin{proof}
	We reduce from \SAT, that is a classical \NP-complete problem. Let $F$ be a boolean formula on $n$ variables in conjunctive normal form $F=C_1 \wedge C_2 \wedge \ldots \wedge C_m$. $C_i$ is a clause being a disjunction of distinct literals, so it can be represented as $C_i=l_{i,1}\vee l_{i,2} \vee \ldots \vee l_{i,k_i}$, where each literal $l_{i,t}$ is either a variable $x_j$ or its negation $\overline{x_j}$ for some $j\in[n]$. 
	
	We show how to, given $F$, construct an instance $(G, p, k)$ of \MHEfull, such that $F$ is satisfiable if and only if $(G, p, k)$ is a yes-instance of \textsc{MHE}. Moreover, $G$ is a threshold graph and the construction can be done in polynomial-time.
	
	Let $F$ be a boolean formula on $n$ variables in CNF, consisting of $m$ clauses. We construct $(G, p, k)$ as follows. 
	

	$G$ will be a threshold graph. So it will consist of two parts: a clique $K$ and an independent set $I$.
	Firstly, we introduce the clique vertices in $G$. For each clause $C_i$ of $F$ we introduce a new vertex $c_i$ in $G$. For each variable $x_j$ of $F$ we introduce $m^2$ new vertices $v_{j,1}, v_{j,2}, \ldots, v_{j, m^2}$ in $G$. We introduce all possible edges between these $m+nm^2$ vertices in $G$ so these vertices form the clique $K$ in the partition of $G$.

	Before we proceed, let us give an intuition of the further construction. Each color we use in $p$ corresponds to a literal in $F$, i.e.\ to an element in $L=\{x_1, x_2, \ldots, x_n, \overline{x_1}, \overline{x_2}, $ $\ldots, \overline{x_n}\}$. Thus, we use $2n$ colors in $p$. For convenience, we use corresponding literals to denote colors instead of the numbers in $[2n]$. We want each clause vertex $c_i$ to be colored with a color corresponding to one of its literals, i.e.\ one of the colors $l_{i,1}, l_{i,2}, \ldots, l_{i,k_i}$ in any optimal coloring. Similarly, we want each variable vertex $v_{j,t}$ corresponding to the variable $x_j$ to be colored with one of the colors corresponding to the literals of $x_j$, i.e.\ either $x_j$ or $\overline{x_j}$. For each vertex $u \in K$, we denote the set of required colors as $\mathcal{L}(u)$, i.e.\ $\mathcal{L}(c_i)=C_i=\{l_{i,1}, l_{i,2}, \ldots, l_{i, k_i}\}$ for clause vertices, and $\mathcal{L}(v_{j,t})=\{x_j, \overline{x_j}\}$ for variable vertices. The purpose of the remaining independent set of $G$ is exactly to ensure that the vertices of the clique are colored with the required colors.

	Our graph is a threshold graph. It means that it is possible to find an order  $u_1, u_2, \dots, u_{|K|}$ of the vertices in $K$ such that $N[u_i] \subseteq N[u_{i+1}]$ for each $i\in[|K|-1]$, i.e. satisfy the condition of Lemma \ref{lemma:threshold_char}. The order we obtain is the following: $u_i=c_i$ for every $i \in [m]$, and $u_{m+jm^2+t}=v_{j+1, t}$ for each $j \in \{0, 1, \ldots, n-1\}$ and each $t \in [m^2]$. The condition of Lemma~\ref{lemma:threshold_char} is satisfied as we step by step add vertices to $I$.
	The $i^{\text{th}}$ step will correspond to the vertex $u_i \in K$. At this step we introduce all neighbours of $u_i$ in $I$ and their colors in the precoloring $p$. For convenience we denote $N(u_i)\cap I$ by $P_i$.
	
	
	At first, we construct  $P_1$ in the following way.
	For each $l \in \mathcal{L}(u_1)$, add exactly $m+nm^2$ vertices to $P_1$ and color them with the color $l$.
	No more vertices are added to $P_1$, so $|P_1|=|\mathcal{L}(u_1)|\cdot (m+nm^2)$.
	Then, for each $i \in [2,m+nm^2]$, we construct $P_i$ by adding new vertices to $P_{i-1}$ and precoloring them. By doing so we satisfy condition $N[u_i]\subseteq N[u_{i+1}]$ for each $i \in [m+nm^2-1]$. 
	The process of this construction is described below  and illustrated in Fig.~\ref{fig:topView}.
	
	\begin{figure}[t]
		\centering{
			\resizebox{	110mm}{!}{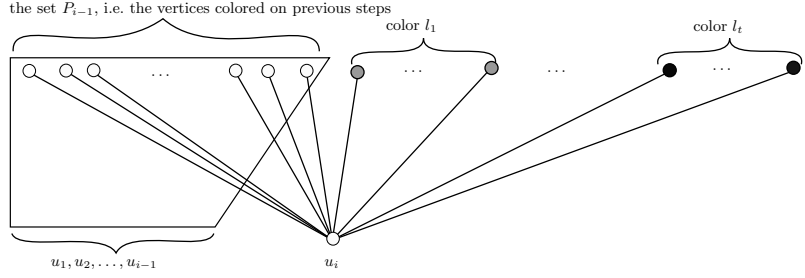}
			\caption{Step $i$. That is addition of the vertex $u_i$ and construction of the set $P_i$. Let $\mathcal{L}(u_i)=\{l_1,l_2, \dots l_t\}$. Vertex $u_i$ is connected to all vertices in $P_{i-1}$. Moreover, for each $j \in \{1,2, \dots t\}$  we introduce $i(m+nm^2)-|\mathcal{N}_p(P_{i-1},l_j)|$ vertices precolored in color $l_j$ and connect them to $u_i$. Recall that all bottom vertices, i.e. $u_q$ for $q \in [i]$, are also pairwise connected (this is not shown in the figure as well as edges from $P_{i-1}$ to $u_1, u_2, \dots, u_{i-1}$). }
			\label{fig:topView}
		}
	\end{figure}
	
	Let $\mathcal{N}_p(P_i, l)$ be the number of vertices in $P_i$ that are precolored with the color $l$, i.e. $\mathcal{N}_p(P_i, l)=\left|\{u \in P_i \mid p(u)=l \}\right|.$
	For each $i$ we require that the vertices in $P_i$ are precolored mostly with required colors for $u_i$, that is, colors in the set $\mathcal{L}(u_i)$. Formally, for every $l \in L$, we require
	\begin{equation*}
		\label{eq:mhe_threshold_constr}
		\tag{*}
		\begin{matrix*}[l]
			\text{$\mathcal{N}_p(P_i, l)=i(m+nm^2)$,} && \text{if $l \in \mathcal{L}(u_i)$},\\
			\text{$\mathcal{N}_p(P_i, l)\le (i-1)\cdot(m+nm^2)$,} && \text{if $l \notin \mathcal{L}(u_i)$.}
		\end{matrix*}
	\end{equation*}
	Note that $P_1$ satisfies \eqref{eq:mhe_threshold_constr}.
	
	Now let $P_{i-1}$ be constructed and satisfy \eqref{eq:mhe_threshold_constr}. We construct $P_i$ that also satisfies the constraint. We start with $P_i=P_{i-1}$. Then for each $l \in \mathcal{L}(u_i)$ we introduce $i(m+nm^2)-\mathcal{N}_p(P_{i-1},l)$ new vertices precolored with color $l$ to $P_i$. For every $l \in \mathcal{L}(u_i)$, $\mathcal{N}_p(P_{i},l)=\mathcal{N}_p(P_{i-1},l)+i(m+nm^2)-\mathcal{N}_p(P_{i-1},l)=i(m+nm^2)$. On the other hand, for every $l \notin \mathcal{L}(u_i)$, $\mathcal{N}_p(P_i, l)=\mathcal{N}_p(P_{i-1},l) \le (i-1)\cdot(m+nm^2)$. Hence, $P_i$ also satisfies \eqref{eq:mhe_threshold_constr}.
	
	The construction of $G$ is finished. Let us remark again that $K$ forms a clique in $G$ and $I=P_{m+nm^2}$ forms an independent set in $G$. 
    We constructed graph in a way that $N[u_i] \subseteq N[u_{i+1}]$ (as $P_i \subseteq P_{i+1}$) for each $i \in [m+nm^2-1]$ where $K=\{u_1, u_2, \ldots, u_{m+nm^2}\}$. Thus, by Lemma \ref{lemma:threshold_char}, $G$ is a threshold graph. Moreover, construction of $G$ is done in polynomial time.
	
	We finally set the number of required happy edges $$k=\left(m+nm^2\right)\cdot\binom{m+nm^2+1}{2}+n\cdot\binom{m^2}{2}+m^3$$
	and argue that $F$ is satisfiable if and only if $(G,p,k)$ is a yes-instance of \textsc{MHE}.
	
	Let $F$ be satisfiable, that is, $F$ has a satisfying assignment $\sigma: x_j \mapsto \{0,1\}$. We construct coloring $c$ of $G$ extending $p$ that yields at least $k$ happy edges as follows.
	
	For each $j \in [n]$ and $t \in [m^2]$, set the color of the vertex $v_{j,t}$ corresponding to the variable $x_j$ with the color corresponding to the literal of $x_j$ that evaluates to $1$ with respect to $\sigma$, i.e.\ $c(v_{j,t})=x_j$, if $\sigma(x_j)=1$, or $c(v_{j,t})=\overline{x_j}$, if $\sigma(x_j)=0$.
	
	For each $i \in [m]$, there is at least one variable satisfying clause $C_i$. In other words, there exists $j \in [n]$, such that either $x_j \in C_i$ and $\sigma(x_j)=1$, or $\overline{x_j} \in C_i$ and $\sigma(x_j)=0$. Choose any such $j$ and color the corresponding clause vertex $c_i$ with the color corresponding to the literal of $x_j$ that evaluates to true. That is, $c(c_i)=x_j$ if $\sigma(x_j)=1$, or $c(c_i)=\overline{x_j}$ if $\sigma(x_j)=0$. There is no any uncolored vertex left, so the construction of $c$ is finished.

	\begin{claimO}
		There are at least $k$ happy edges in $G$ with respect to $c$. 
	\end{claimO}
	\begin{claimproofO}
		Consider edges between $K$ and $I$ in $G$. Observe that for each $u_i \in K$, $c(u_i) \in \mathcal{L}(u_i)$, that is, each variable vertex is colored with the color corresponding to one of its literals, and each clause vertex is colored with the color corresponding to one of the literals it contains. Each $u_i \in K$ is incident to exactly $\mathcal{N}_p(P_i, c(u_i))$ happy edges that has the other endpoint in $I$. By construction of $G$, any $P_i$ satisfies \eqref{eq:mhe_threshold_constr} and $c(u_i) \in \mathcal{L}(u_i)$, hence there are exactly $i(m+nm^2)$ happy edges between $u_i$ and $I$ in $G$ with respect to $c$.
		
		In total, there are exactly
		$$\sum_{i=1}^{m+nm^2}i(m+nm^2)=(m+nm^2)\cdot \sum_{i=1}^{m+nm^2}i=(m+nm^2)\cdot\binom{m+nm^2+1}{2}$$
		happy edges between $K$ and $I$ with respect to $c$.
		
		For each $j \in [n]$, variable vertices $\{v_{j,1}, v_{j,2}, \ldots, v_{j,m^2}\}$ share the same color, and form a clique in $G$. Thus, for each $j \in [n]$, there are $\binom{m^2}{2}$ happy edges between vertices of type $v_{j,t}$. In total over all variable vertices, there are $n \cdot \binom{m^2}{2}$ such happy edges with respect to $c$.
		
		Consider now edges between clause vertices and variable vertices. For each $i \in [m]$, clause vertex $c_i$ is colored with the color corresponding to a literal that evaluates to $1$ with respect to $\sigma$. This literal corresponds to some variable, say $x_j$. All variable vertices corresponding to $x_j$ are also colored with the literal of $x_j$ that evaluates to $1$ with respect to $\sigma$. Hence, for each $t \in [m^2]$, $c(v_{j,t})=c(c_i)$. On the other hand, $c(v_{j',t})\neq c(c_i)$ for any $j' \neq j$, since $c(v_{j',t})$ corresponds to a literal of the variable $x_{j'}$. Thus, there are exactly $m^2$ happy edges between $c_i$ and the variable vertices $v_{j,t}$ in $G$ with respect to $c$. In total over all clause vertices, there are exactly $m^3$ such happy edges.
		
		Considered types of edges are distinct and cover all edges of $G$. The number of happy edges among them sums up to $k$.
	\end{claimproofO}

	Hence, we showed that if $F$ is satisfiable, then $(G, p, k)$ is a yes-instance of \textsc{MHE}. We now give a proof in the other direction.
	
	Let $c$ be a coloring of $G$ extending $p$ such that at least $k$ edges of $G$ are happy with respect to $c$. We assume that $c$ is optimal, i.e.\ it yields the maximum number of happy edges in $G$. We make the following claims and then show how to construct a satisfying assignment $\sigma$ of $F$.
	
	\begin{claimO}\label{claim:mhe_thr_colored_required}
		In any optimal coloring $c$ of $G$ extending $p$, $c(u_i) \in \mathcal{L}(u_i)$ for every $u_i \in K$.
	\end{claimO}
	\begin{claimproofO}
		Suppose that $c$ is an optimal coloring of $G$, but $c(u_i) \notin \mathcal{L}(u_i)$ for some $u_i \in K$. There are exactly $\mathcal{N}_p(P_i, c(u_i))$ happy edges between $u_i$ and $I$. On the other hand, $|K|=m+nm^2$, hence $u_i$ is adjacent to at most $m+nm^2-1$ vertices of color $c(u_i)$ in $K$.
		
		By \eqref{eq:mhe_threshold_constr}, $\mathcal{N}_p(P_i, c(u_i))\le (i-1)\cdot(m+nm^2)$, hence $u_i$ is incident to at most $|K|-1+\mathcal{N}_p(P_i, c(u_i))\le i(m+nm^2)-1$ happy edges. But if one picks any color $l \in \mathcal{L}(u_i)$ and puts $c(u_i)=l$, $u_i$ becomes adjacent to at least $\mathcal{N}_p(P_i, l)=i(m+nm^2)$ happy edges, and the total number of happy edges in $G$ with respect to $c$ increases. This contradicts the optimality of $c$.
	\end{claimproofO}

	\begin{claim}\label{claim:mhe_thr_variables_same}
		In any optimal coloring $c$ of $G$ extending $p$, all variable vertices corresponding to the same variable are colored with the same color.  Formally, $c(v_{j,t_1})=c(v_{j, t_2})$ for every $j \in [n]$ and $t_1, t_2 \in [m^2]$.
	\end{claim}
	\begin{claimproof}
		Suppose that $c$ is an optimal coloring extending $p$, but $c(v_{j, t_1})\neq c(v_{j, t_2})$ for some $j \in [n]$, $t_1, t_2 \in [m^2]$. By Claim \ref{claim:mhe_thr_colored_required}, $c(v_{j, t_1})$ and $c(v_{j, t_2})$ are distinct literals of variable $x_j$, and $v_{j,t_1}=u_{m+(j-1)m^2+t_1}$ and $v_{j,t_2}=u_{m+(j-1)m^2+t_2}$ are incident to exactly $(m+(j-1)\cdot m^2+t_1) \cdot (m+nm^2)$ and $(m+(j-1)\cdot m^2+t_2) \cdot (m+nm^2)$ happy edges going in $I$ respectively according to \eqref{eq:mhe_threshold_constr}.
		
		Let $h_1$ and $h_2$ be the number of vertices in $K$ that are colored with colors $c(v_{j,t_1})$ and $c(v_{j,t_2})$, respectively. Thus, $v_{j,t_1}$ and $v_{j,t_2}$ are incident to exactly $h_1-1$ and $h_2-1$ happy edges in $G[K]$, respectively. Note that the edge between $v_{j,t_1}$ and $v_{j,t_2}$ is not happy.
		
		Without loss of generality, $h_1 \ge h_2$. Change the color of $v_{j,t_2}$ in $c$ to $c(v_{j,t_1})$. Since $c(v_{j,t_2})$ is still a literal of $x_j$, hence $c(v_{j,t_2}) \in \mathcal{L}(v_{j,t_2})$, the number of happy edges connecting $v_{j,t_2}$ and $I$ does not change, even though the set of such happy edges becomes different. Consider edges in $G[K]$. $v_{j,t_2}$ is now adjacent to $h_1$ neighbours of the same color, as the edge between $v_{j,t_1}$ and $v_{j,t_2}$ also becomes happy. Since $h_1 > h_2-1$, we have increased the total number of happy edges in $G$ with respect to $c$. This contradicts the optimality of $c$.
	\end{claimproof}

	We now use the above claims to construct $\sigma$ from an optimal coloring $c$ yielding at least $k$ happy edges. By Claim \ref{claim:mhe_thr_colored_required}, there are exactly $(m+nm^2)\cdot(1+2+\ldots+(m+nm^2))=(m+nm^2)\cdot\binom{m+nm^2+1}{2}$ happy edges between $K$ and $I$ with respect to $c$. By Claim \ref{claim:mhe_thr_variables_same}, there are exactly $n\cdot\binom{m^2}{2}$ happy edges between all variable vertices. There are exactly $m$ clause vertices in $G$, hence there are at most $\binom{m}{2}$ happy edges between all clause vertices. The only edges left are the edges between clause and variable vertices, hence there are at least
	$$k-(m+nm^2)\cdot\binom{m+nm^2+1}{2}-n\cdot\binom{m^2}{2}-\binom{m}{2}=m^3-\binom{m}{2}$$
	happy edges between clause and variable vertices.
	
	Construct $\sigma$ according to the colors of variable vertices, so that the literal corresponding to $c(v_{j,t})$ evaluates to $1$ with respect to $\sigma$. Formally, for each $j\in[n]$, $\sigma(x_j)=1$ if $c(v_{j,1})=x_j$, and $\sigma(x_j)=0$ if $c(v_{j,1})=\overline{x_j}$. We now argue that each clause $C_i \in F$ contains a literal that evaluates to $1$ with respect to $\sigma$, and that this literal is $c(c_i)$.
	
	Suppose that it is not true, and there is a clause $C_i$ so that $c(c_i)$ is a literal that evaluates to $0$ with respect to $\sigma$. By construction of $\sigma$, there are no happy edges between $c_i$ and the variable vertices.
	$c_i$ corresponds to a literal evaluating to $0$, but all colors of variable vertices are literals that evaluates to $1$ with respect to $\sigma$.
	Moreover, any other clause vertex $c_{i'}$ is adjacent to either $0$ or $m^2$ variable vertices of color $c(c_{i'})$.
	For each literal, there are either $0$ or $m^2$ variable vertices colored correspondingly to this literal.
	
	There are exactly $m-1$ clause vertices apart from $c_i$, hence at most $(m-1)\cdot m^2$ edges between clause vertices and variable vertices are happy with respect to $c$. But $(m-1) \cdot m^2=m^3-m^2<m^3-\binom{m}{2}$ for any $m>0$, a contradiction. Thus, each clause $C_i$ contains a literal that evaluates to $1$ with respect to $\sigma$, i.e.\ $\sigma$ is a satisfying assignment of $F$. We proved that if $(G, p, k)$ is a yes-instance of \textsc{MHE}, then $F$ is satisfiable.
	The proof is complete.
\end{proof}

\begin{corollary}\label{cor:mhe_cliquewidth_xp}
	There is no \XP-algorithm for \MHEfull~parameterized by clique-width, unless $\P=\NP$.
\end{corollary}
\begin{proof}
	Suppose there is an \XP-algorithm for \MHEfull~parameterized by clique-width, i.e.\ there is an algorithm with running time $n^{f(\text{cw})}$ for some function $f$ for \MHE. Threshold graphs are a subclass of cographs \cite{zbMATH00803309}, that is, graphs of clique-width at most two \cite{Courcelle2000}. Hence, \MHE~on threshold graphs can be solved in $n^{f(2)}=n^{\O(1)}$. Then, by Theorem \ref{thm:mhe_np_hard}, problem that is solvable in polynomial time is \NP-hard, hence $\P=\NP$.
\end{proof}

We have shown that \MHE~parameterized by clique-width alone is hard.
Following known results on the existence of  $\Ostar{\ell^{\O(\operatorname{pw})}}$ and $\Ostar{\ell^{\O(\operatorname{tw})}}$
running time algorithms for both \MHV~and \MHE~parameterized by pathwidth or treewidth combined with the number of colors $\ell$ \cite{Agrawal2018,Misra2018,Aravind2016}, it is reasonable to ask the complexity of \MHE~parameterized by $\text{cw}+\ell$.
We now show that \MHE~parameterized by $\text{cw}+\ell$ admits an \XP-algorithm.

\begin{theorem}\label{thm:mhe_cliquewidth}
	There is an algorithm for \MHEfull~with $n^{\O(\ell \cdot \operatorname{cw})}$ running time, if a $\operatorname{cw}$-expression of $G$ is given.
\end{theorem}
\begin{proof}
	The algorithm is by standard dynamic programming on a given $w$-expression $\Psi$ of $G$.
	We assume that $\Psi$ is a nice $w$-expression of $G$, i.e.\ no edge is introduced twice in $\Psi$.
	For each subexpression $\Phi$ of $\Psi$,
	$$OPT(\Phi, n_{1,1}, n_{1,2}, \ldots, n_{1,\ell}, n_{2,1}, n_{2,2},\ldots, n_{w,\ell-1}, n_{w,\ell})$$
	denotes the maximum number of happy edges that can be obtained in $G_\Phi$ simultaneously with respect to a coloring such that the number of vertices with a label $i$ in $G_\Phi$ that are colored with a color $a$ in $G_\Phi$ is exactly $n_{i,a}$.
	Formally,
	$$OPT(\Phi,n_{1,1},\ldots,n_{w,\ell})=\max\left\{
		|\mathcal{E}(G_\Phi,c)| \left|
		\begin{matrix}
			c: V(G_\Phi) \to [\ell],\\
			\forall i \in [w], a \in [\ell]: \;\;|c^{-1}(a)\cap V_i(\Phi)|=n_{i,a}  \\
		\end{matrix}
		\right.
	\right\},$$
	where $\mathcal{E}(G_\Phi, c)$ is the set of edges that are happy in $G_\Phi$ with respect to $c$.
	If there are no colorings corresponding to a cell $OPT(\Phi, n_{1,1},\ldots, n_{w,\ell})$, we put its value equal to $-\infty$.
	
	The algorithm computes the values of $OPT$ in a bottom-up approach, starting from the simplest subexpressions of $\Psi$ up to $\Psi$ itself.
	Thus, when the algorithm starts computing the values of $OPT(\Phi,\cdot)$ for a subexpression $\Phi$ of $\Psi$, it has all values of $OPT$ computed for each subexpression of $\Phi$.
	There are four possible cases of computing values of $OPT(\Phi,\cdot)$ depending on the topmost operator in $\Phi$.
	
	\begin{enumerate}
		\item $\Phi = i(v)$.
		Since $G_\Phi$ contains a single vertex with label $i$ and no edges, it is enough to iterate over all possible colors of this vertex.
		If $v$ is not precolored, for each color $a \in [\ell]$ put $OPT(\Phi, 0, \ldots, 0, n_{i,a}=1, 0, \ldots, 0)=0.$
		Otherwise, the color of $v$ can only be $p(v)$, so do this only for $a=p(v)$.
		Thus, exactly $\ell$ values (or exactly one value) of $OPT(\Phi,\cdot)$ are put equal to $0$, and all other values should equal $-\infty$ by the definition of $OPT$.
		\item\label{enu:mhe_cw_case_disj_union} $\Phi = \Phi' \oplus \Phi''$.
		Consider a cell $OPT(\Phi, n_{1,1}, \ldots, n_{w,\ell})$.
		Any coloring $c$ corresponding to this cell is split uniquely in the two colorings $c'=c|_{V(G_{\Phi'})}$ and $c''=c|_{V(G_{\Phi''})}$ of $G_{\Phi'}$ and $G_{\Phi''}$ respectively.
		In its order, these colorings correspond to cells $OPT(\Phi', n'_{1,1}, \ldots, n'_{w,\ell})$ and $OPT(\Phi'', n''_{1,1}, \ldots, n''_{w,\ell})$, where $n'_{i,a}$ and $n''_{i,a}$ are unique for each choice of $i \in [w], a \in [\ell]$.
		As $G_\Phi$ is the disjoint union of $G_{\Phi'}$ and $G_{\Phi''}$, the number of happy edges in $G_\Phi$ with respect to $c$ can be found as a sum of happy edges with respect to $c'$ and $c''$ in the corresponding graphs.
		Hence,
		\begin{multline}OPT(\Phi,n_{1,1},\ldots, n_{w,\ell})=\\\max\limits_{n'_{i,a}+n''_{i,a}=n_{i,a}} \left\{OPT(\Phi',n'_{1,1},\ldots, n'_{w,\ell})+OPT(\Phi'',n''_{1,1},\ldots,n''_{w,\ell})\right\}.\end{multline}
		\item $\Phi = \rho_{i \to j} \Phi'$.
		Consider again a coloring $c$ corresponding to a cell $OPT(\Phi, n_{1,1},\ldots, n_{w,\ell})$.
		Note that $\Phi$ contains no vertices with label $i$, so $n_{i,a}=0$ for each $a \in [\ell]$.
		Moreover, $c$ is a coloring of $G_{\Phi'}$, thus it corresponds to the unique cell $OPT(\Phi, n'_{1,1},\ldots, n'_{w,\ell})$, where $n'_{i,a}+n'_{j,a}=n_{j,a}$ for each $a \in [\ell]$ and $n'_{k,a}=n_{k,a}$ for each choice of label $k$ distinct from $i$ and $j$, and for each color $a \in [\ell]$.
		The number of happy edges in $G_{\Phi}$ with respect to $c$ is the same as that in $G_{\Phi'}$.
		Hence,
		\begin{multline}OPT(\Phi,n_{1,1}, \ldots, n_{w,\ell})=\\\left\{\begin{array}{l}
		-\infty, \text{ if } \exists a \in [\ell]: n_{i, a}\neq 0,\\
		\max
		\left\{
		OPT(\Phi',n'_{1,1},\ldots,n'_{w,\ell})
			\left|
			\begin{matrix}
				\forall a \in [\ell]: \; n'_{i,a}+n'_{j,a}=n_{j,a}\\
				\forall k \in [w] \setminus \{i,j\}, a \in [\ell]: \; n'_{k,a}=n_{k,a}\\
			\end{matrix}
			\right.
		\right\}
		\end{array}
		\right..\end{multline}

		\item $\Phi = \eta_{i,j} \Phi'$.
		This is the only case where edges are introduced.
		Any coloring $c$ of $G_\Phi$ is a coloring of $G_\Phi'$.
		Moreover, if $c$ corresponds to $OPT(\Phi,n_{1,1}, \ldots, n_{w,\ell})$ then, clearly, $c$ corresponds to $OPT(\Phi',n_{1,1}, \ldots, n_{w,\ell})$ as well.
		Thus, one shall only compute the number of newly-introduced edges that are happy with respect to $c$.
		As $\Psi$ is a nice $w$-expression, each edge between vertices with label $i$ and vertices with label $j$ is newly-introduced.
		Each of such happy edge should connect a vertex with the label $i$ and a vertex with the label $j$ that are colored with the same color $a$ for some $a$.
		The number of such edges for a fixed $a$ is $n_{i,a} \cdot n_{j,a}$.
		Hence,
		$OPT(\Phi,n_{1,1},\ldots,n_{w,\ell})=OPT(\Phi',n_{1,1},\ldots,n_{w,\ell})+\sum_{a=1}^\ell n_{i,a}\cdot n_{j,a}.$
	\end{enumerate}

	The description of all possible cases for $\Phi$ and corresponding recurrence relations is finished.
	Note that there are at most $|\Psi| \cdot n^{\ell \cdot w}$ cells in the $OPT$ table, and each of them is computed in $\O(n^{\ell \cdot w})$ time (computation in the case of disjoint union and the case of relabelling takes the most time) by the algorithm.
	Thus, the whole computation of $OPT$ takes $\O(|\Psi|\cdot n^{2\ell \cdot w})$ running time.
	Clearly, the maximum number of happy edges that can be obtained in $G$ simultaneously equals $\max_{n_{1,1},\ldots,n_{w,\ell}} OPT(\Psi,n_{1,1},\ldots,n_{w,\ell}),$
	which is found in $\O(n^{\ell \cdot w})$ time.
	This finishes the proof.
\end{proof}

\begin{corollary}
	\MHEfull~parameterized by ${\normalfont\text{cw}}+\ell$ admits an \XP-algorithm.
\end{corollary}

The fixed-parameter tractability of \MHE~with respect to $\operatorname{cw}+\ell$ remains unknown though.
We note that Theorem \ref{thm:mhe_cliquewidth} does not imply that no \FPT-algorithm exists for \MHE~parameterized by $\text{cw}+\ell$ under $\P\neq \NP$.
But it at least implies that no algorithm with running time $\Ostar{\poly(\ell)^{f(\operatorname{cw})}}$ exists for \MHE, unless $\P=\NP$.
We leave the \FPT-membership of \MHE~parameterized by $\text{cw}+\ell$ as an open question.
	\section{Maximum Happy Vertices}

We start this section by answering the complexity of \MHVfull~parameterized by $\text{cw}+\ell$.
We note that \MHV~is \W[2]-hard when parameterized by the clique-width of the input graph alone \cite{bliznets2019on}.
In contrast to this, we show that \MHV~is in \FPT~if the clique-width parameter is extended by the number of colors $\ell$.

\begin{theoremO}\label{thm:mhv_cliquewidth}
	\MHVfull~can be solved in $(\ell+1)^{\O(\operatorname{w})}\cdot n^{\O(1)}$ running time, if a $\operatorname{w}$-expression of $G$ is given.
\end{theoremO}
\begin{proofO}
	Given a $w$-expression $\Psi$ of $G$, we solve $(G,p,k)$ with the following dynamic programming.
	$$OPT(\Phi, col_1, col_2, \ldots, col_w, out_1, out_2, \ldots, out_w),$$ where $col_i, out_i \in [\ell]\cup\{0\}$, and $\Phi$ is a subexpression of $\Psi$. For convenience, we may also refer to this value as $OPT(\Phi, col, out)$, where $col=(col_1, col_2, \ldots, col_w)$ and $out=(out_1, out_2, \ldots, out_w)$, $col, out \in ([\ell]\cup \{0\})^w$. Let $G_{\Phi}$ be a graph expressed by ${\Phi}$, and $V(G_{\Phi})=V_1\sqcup V_2\sqcup \ldots \sqcup V_k$, where $V_i$ is the set of all vertices in $G_{\Phi}$ with label $i$ in $\Phi$. Then $OPT(\Phi, col, out)$ denotes the number of \emph{special} (to be formally defined below) happy vertices in $V(G_{\Phi})$ maximized over all colorings $c$ of $(G,p)$ such that
	\begin{enumerate}
		\item $c$ is a full coloring of $G$ extending $p$;
		\item For each $i$, 
		\begin{itemize}
			\item if $col_i=0$, then either $V_i$ is empty or $|c(V_i)|\ge 2$;  (here and below, $c(S)=\bigcup_{v\in S}\{c(v)\}$ for any $S \subseteq V(G)$)
			\item otherwise, $c(V_i)=\{col_i\}$;
		\end{itemize}
		\item For each $i$,
		\begin{itemize}
			\item if $out_i\neq 0$, then $c(N_G(V_i) \setminus N_{G_{\Phi}}(V_i))=\{out_i\}$.
		\end{itemize}
	\end{enumerate}

	We denote the set of all colorings that satisfy the conditions above by $\mathcal{C}(\Phi, col, out)$. To explain the purpose of the $out$ values, we suggest the following useful observation.
	
	\begin{observation}\label{obs:mhv_cw_outside}
		For each subexpression $\Phi$ of $\Psi$, each $i \in [w]$ and each $v \in V_i$, $$N_G(v)\setminus N_{G_{\Phi}}(v)=N_G(V_i)\setminus N_{G_{\Phi}}(V_i).$$ In other words, vertices with the same label have the same set of neighbours apart from neighbours in $G_{\Phi}$.
	\end{observation}

	With this observation, it is easy to see that, if specified, $out_i\neq 0$ denotes the color of the neighbours of the vertices with label $i$, apart from their neighbours in $G_{\Phi}$.
	In $OPT(\Phi,col,out)$ all happy vertices are counted, except for vertices in $V_i$ such that $out_i = 0$ and $N_G(V_i)\neq N_{G_\Phi}(V_i)$ (that is, vertices with label $i$ having at least one neighbour outside $G_\Phi$).
	That is, $$OPT(\Phi, col, out)=\max\limits_{c \in \mathcal{C}(\Phi, col, out)}\left|\mathcal{H}(G,c) \cap \bigcup\limits_{i=1}^w \mathcal{V}_i(\Phi, out_i)\right|,$$
	where $\mathcal{V}_i(\Phi, out_i)=V_i$ if [$out_i\neq 0$ or $N_G(V_i)=N_{G_\Phi}(V_i)$], and $\mathcal{V}_i(\Phi, out_i)=\emptyset$ otherwise.  We recall that  by $\mathcal{H}(G,c)$ we denote the set of all vertices in $G$ that are happy with respect to $c$.  For compactness, denote $\mathcal{V}(\Phi, out)=\bigcup\limits_{i=1}^{w}\mathcal{V}_i(\Phi, out_i)$.

	Note that the conditions impose that if $V_i$ is empty, then $col_i=0$; and if $N_G(V_i)=N_{G_\Phi}(V_i)$, then $out_i=0$. Therefore, for some values of $\Phi$, $col$ and $out$, there may be no corresponding colorings $c$ of $(G,p)$, i.e.\ $\mathcal{C}(\Phi, col, out)=\emptyset$. For such cases, we put $OPT(\Phi, col, out)=-\infty$. Technically, $-\infty$ is a special value with a property that $x+(-\infty)=(-\infty)+x=-\infty$ and $\max\{x,-\infty\}=\max\{-\infty, x\}=x$ for all possible values of $x$.
	
	To avoid trivial cases when $\mathcal{C}(\Phi, col, out)=\emptyset$, we introduce the notion of \emph{good triples}.
	It has a property that if a triple $(\Phi, col, out)$ is not a good triple, then $\mathcal{C}(\Phi, col, out)=\emptyset$.
	This does not work in the other direction though, and $\mathcal{C}(\Phi, col, out)=\emptyset$ may hold for a good triple $(\Phi, col, out)$.

	\begin{definition}
		We say that a triple $(\Phi, col, out)$ is a \emph{good triple}, if it satisfies the following conditions:
		\begin{enumerate}
			\item For each $i \in [w]$, if $V_i=\emptyset$, then $col_i=0$;
			\item For each $i \in [w]$, if $N_G(V_i)\setminus N_{G_{\Phi}}(V_i)=\emptyset$, then $out_i=0$;
			\item For each $i, j \in [w]$ such that $i \neq j$, $out_i\neq 0$, $V_j \neq \emptyset$, if $V_j \subseteq N_G(V_i)\setminus N_{G_\Phi}(V_i)$, then $col_j=out_i$.
			\item For each $i, j \in [w]$, such that $out_i \neq 0$ and $out_j \neq 0$, if $(N_G(V_i)\setminus N_{G_\Phi}(V_i)) \cap (N_G(V_j)\setminus N_{G_\Phi}(V_j)) \neq \emptyset$, then $out_i = out_j$.
		\end{enumerate}
	\end{definition}
	The first two conditions for a good triple were discussed slightly above. If $(\Phi, col, out)$ does not satisfy these two conditions, then $\mathcal{C}(\Phi, col, out)=\emptyset$.
	The third condition of a good triple handles the case when vertices with label $j$ are neighbours of the vertices with label $i$ in $G$, but not in $G_\Phi$ yet.
	If the color of the outer neighbourhood of the vertices with label $i$ is specified, i.e.\ $out_i\neq0$, then all vertices with label $j$ in $G_\Phi$ should share the same color $out_i$, i.e.\ $col_j=out_i$.
	Obviously, if a triple does not satisfy this condition, there is no colorings corresponding to that triple. The fourth condition ensures that for any two labels sharing an outer neighbour, if the colors of outer neighbours are fixed for both labels, these fixed colors should be the same. So we have:
	\begin{claim}
		Any triple $(\Phi, col, out)$ with $\mathcal{C}(\Phi, col, out)\neq \emptyset$ is a good triple.
	\end{claim}

	From now on, we work just with good triples. That is, we do not necessarily exclude all triples $(\Phi, col, out)$ with $\mathcal{C}(\Phi, col, out)=\emptyset$, but just some of them. More importantly, we do not exclude any triple with $\mathcal{C}(\Phi, col, out)\neq\emptyset$ from our consideration.

	We also note that for each fixed subexpression $\Phi$ and each coloring $c$ of $(G,p)$, there is at least one correct choice of the corresponding values of $col$ and $out$, so that $c \in \mathcal{C}(\Phi, col, out)$.
	
	Obviously, the maximum number of happy vertices that can be obtained in $(G,p)$ can be found as a maximum value of $OPT(\Psi, col, out)$ over all possible values of $col$ and $out$. We now show how we calculate the value of $OPT(\Phi, col, out)$ for each possible choice of the subexpression $\Phi$ of $\Psi$, $col$ and $out$. We do that in a bottom-up manner, starting with the smallest subexpressions of $\Phi$. In fact, when we are to calculate the values of $OPT$ for a subexpression $\Phi$, we have all values of $OPT$ for all proper subexpressions of $\Phi$ calculated. Now fix $\Phi$, $col$ and $out$, for which $(\Phi, col, out)$ is a good triple, and consider the last operator in $\Phi$. Note that if $(\Phi, col, out)$ is not a good triple, then $OPT(\Phi, col, out)$ just equals $-\infty$.
	
	\begin{enumerate}
		\item $\Phi=i(v)$. Then $G_{\Phi}$ is a subgraph of $G$ consisting of a single vertex $v$. Take any $c \in \mathcal{C}(\Phi, col, out)$. Note that $col_i \neq 0$, as $V_i=\{v\}$ and $|c(V_i)|=1$, so $c(v)=col_i$. If $v$ is a precolored vertex in $(G,p)$, then $c(v)$ should equal $p(v)$. Hence, if $v$ is precolored and $col_i \neq p(v)$, then $\mathcal{C}(\Phi, col, out)=\emptyset$, and we put $OPT(\Phi, col, out)=-\infty$.
		
		For each $j\neq i$, $V_j=\emptyset$, so $col_j=out_j=0$ for all such $j$. Since $N_{G_{\Phi}}(v)=\emptyset$, $c(N_G(v))=c(N_{G}(V_i)\setminus N_{G_{\Phi}}(V_i))$. Thus, if $out_i=0$, then either $N_G(v)$ is empty and $v$ is isolated and happy; or $v$ should not be counted in $OPT(\Phi, col, out)$, even if it is happy. If $out_i\neq 0$, $v$ is not isolated in $G$ and all its neighbours are colored with the color $out_i$ by $c$, hence $v$ is happy if and only if $col_i=out_i$. We get that $OPT(\Phi, col, out)=1$ if [$N_G(v)=\emptyset$ or $col_i=out_i$], and $OPT(\Phi, col, out)=0$ otherwise.
		\item $\Phi=\Phi' \oplus  \Phi''$. Let $V(G_{\Phi'})=V'_1 \sqcup V'_2 \sqcup \ldots \sqcup V'_w$ and $V(G_{\Phi''})=V''_1 \sqcup V''_2 \sqcup \ldots \sqcup V''_w$ in the same way as for $V(G_{\Phi})$. In particular, $V_i=V'_i \sqcup V''_i$ for each $i$.
		
		Also, similarly to Observation \ref{obs:mhv_cw_outside}, $N_G(V_i)\setminus N_{G_{\Phi}}(V_i)=N_G(V'_i)\setminus N_{G_{\Phi'}}(V'_i)$ if $V'_i$ is not empty and $N_G(V_i)\setminus N_{G_{\Phi}}(V_i)=N_G(V''_i)\setminus N_{G_{\Phi''}}(V''_i)$ if $V''_i$ is not empty for each $i$.
		
		Take any coloring $c\in \mathcal{C}(\Phi, col, out)$. Let $out'_i=out_i$ if $V'_i$ is not empty and $out'_i=0$ otherwise. Define $out''$ analogously. Then $c \in \mathcal{C}(\Phi', col', out')$ and $c \in \mathcal{C}(\Phi'', col'', out'')$ for some choice of the values of $col'$ and $col''$. Consider a label $i$. If $col_i\neq 0$, then $col'_i=col''_i=col_i$ (with an exception that if $V'_i$ or $V''_i$ is empty, $col'_i$ or $col''_i$ should equal $0$ respectively). It is left to consider $col_i=0$. If $V_i$ is empty, then both $V'_i$ and $V''_i$ are empty, so $col'_i=col''_i=0$ as well. In the other case, $|c(V_i)|\ge 2$. Then it is either $c(V'_i)=\{col'_i\}$ and $c(V''_i)=\{col''_i\}$ and $col'_i \neq col''_i$, or at least one of $c(V'_i)$ and $c(V''_i)$ has size at least two. That is, $V'_i\neq \emptyset$ and $col'_i=0$, or $V''_i\neq \emptyset$ and $col''_i=0$. Denote by $P_i(col_i)$ the set of all such appropriate pairs $(col'_i, col''_i)$ for $col_i$. That is, $(col'_i, col''_i) \in P_i(col_i)$ if and only if $(col'_i, col''_i)$ satisfies
		\begin{enumerate}
			\item If $V'_i=\emptyset$, then $col'_i=0$;
			\item If $V''_i=\emptyset$, then $col''_i=0$;
			\item If $col_i\neq 0$, then
			\begin{enumerate}
				\item $col'_i=col_i$ if $V'_i\neq \emptyset$; and
				\item $col''_i=col_i$ if $V''_i \neq \emptyset$;
			\end{enumerate}
			\item If $col_i=0$, then either
			\begin{enumerate}
				\item $V'_i \neq \emptyset$ and $V''_i\neq\emptyset$ and $col'_i\neq col''_i$; or
				\item $V'_i \neq \emptyset$ and $col'_i=0$; or
				\item $V''_i \neq \emptyset$ and $col''_i=0$.
			\end{enumerate}
		\end{enumerate}
	
		We formulate the discussion above in the following claim.
		\begin{claim}\label{claim5}
			Let $P(col)=\{(col', col'') \;\mid\; \forall i \in [w]: (col'_i, col''_i) \in P_i(col_i)\}$. Let $out'_i=out_i$ if $V'_i\neq \emptyset$, and $out'_i=0$ otherwise. Define $out''_i$ in the same way for $V''_i$. Then $$\mathcal{C}(\Phi, col, out)=\bigcup\limits_{(col', col'') \in P(col)}\mathcal{C}(\Phi', col', out')\cap \mathcal{C}(\Phi'', col'', out'').$$
		\end{claim}
	
		We shall now formulate the main lemma for this case.
	
		\begin{lemma}\label{lemma:mhv_cw_disj_main_equ}
			Let $out'_i=out_i$ if $V'_i\neq \emptyset$, and $out'_i=0$ otherwise. Define $out''_i$ in the same way for $V''_i$.
			Then
		$$\label{equ:cw_disj_union_mhv}
			OPT(\Phi, col, out)=\max\limits_{(col', col'') \in P(col)} \left\{OPT(\Phi', col', out')+OPT(\Phi'', col'', out'') \right\}.
		$$
		\end{lemma}
	
		Note that this lemma does not immediately follow from the definition of $P(col)$.
		The maximum number of happy vertices $OPT(\Phi', col', out')$ in $V(G_{\Phi'})$ is achieved with some coloring $c' \in \mathcal{C}(\Phi', col', out')$, and the maximum number of happy vertices in $V(G_{\Phi''})$ is achieved with some coloring $c'' \in \mathcal{C}(\Phi'', col'', out'')$. But in general, $c'$ and $c''$ are different colorings of $(G,p)$.
		Therefore, we need a mechanism to combine two different colorings that agree with $(\Phi, col, out)$ in one coloring that preserves all happy vertices in both $\mathcal{V}(\Phi', out')$ and $\mathcal{V}(\Phi'', out'')$. We proceed with the following claim, which implies Lemma \ref{lemma:mhv_cw_disj_main_equ}.
		
		\begin{claimO}\label{claim:mhv_cliquewidth_merge_colorings}
			Let $out'_i=out_i$ if $V'_i\neq \emptyset$, and $out'_i=0$ otherwise. Define $out''_i$ in the same way for $V''_i$.
			Let $(col', col'') \in P(col)$, and let $c' \in \mathcal{C}(\Phi', col', out')$, $c''\in \mathcal{C}(\Phi'', col'', out'')$.
			Let $c$ be a coloring of $(G,p)$ defined as
			$$c(v) = \left\{
			\begin{matrix*}[l]
				c''(v), \text{ if $v \in V(G_{\Phi''})\cup N_G(\mathcal{V}(\Phi'', out''))$};\\
				c'(v), \text{ otherwise}.
			\end{matrix*}\right.$$
			Then  (i) $c \in \mathcal{C}(\Phi', col', out')\cap \mathcal{C}(\Phi'', col'', out'')$,
			(ii) $c \in \mathcal{C}(\Phi,col,out)$, (iii) $\mathcal{H}(G,c)\supseteq (\mathcal{H}(G,c')\cap \mathcal{V}(\Phi', out'))$  and $\mathcal{H}(G,c)\supseteq (\mathcal{H}(G,c'')\cap \mathcal{V}(\Phi'', out''))$.
		\end{claimO}
		\begin{claimproofO}
			(i) We first show that $c$ is contained in both $\mathcal{C}(\Phi', col', out')$ and $\mathcal{C}(\Phi'', col'', out'')$. Consider $\mathcal{C}(\Phi'', col'', out'')$. $c$ preserves all colors of the vertices of $G_{\Phi''}$ in $c''$, so the restrictions imposed by $col''$ are satisfied in $c$. Moreover, for each $i \in [w]$ with $out_i'' \neq 0$, $c$ also preserves the colors of the outer neighbours of vertices with label $i$. Thus, the restrictions imposed by $out''$ are satisfied in $c$ as well, and $c \in \mathcal{C}(\Phi'', col'', out'')$. Now consider $\mathcal{C}(\Phi', col', out')$, and suppose that $c \notin \mathcal{C}(\Phi', col', out')$. Then, by definition of $\mathcal{C}$, there is a vertex $v$ so that $c(v)$ does not satisfy the restrictions imposed by $col'$ and $out'$. Note that this may only happen if $c(v)=c''(v)$.
			
			If the unsatisfied restrictions are imposed by $col'$, then $v \in V_i'$ for some $i$, $col_i' \neq 0$ and $c''(v)\neq col_i'$. Since $c(v)=c''(v)$, $v \in V(G_{\Phi''})\cup N_G(\mathcal{V}(\Phi'', out''))$. As $V(G_{\Phi'})$ and $V(G_{\Phi''})$ are disjoint, $v \in N_G(\mathcal{V}(\Phi'', out''))$. That is, $v$ is a neighbour of a vertex $u \in \mathcal{V}_j(\Phi'', out'')$ for some $j$.
			Hence, by definition of $\mathcal{V}_j$, $out_j''\neq 0$, and $c''(v)=out_j''$.
			Here we have $v \in N_G(V''_j)\setminus N_{G_{\Phi''}}(V''_j)$.
			Note that the edge between $u$ and $v$ is outside of $G_\Phi$, so $V_i \subseteq N_G(V_j)\setminus N_{G_\Phi}(V_j)$. Since $out_j''=out_j$, $out_j\neq 0$, and as $(\Phi, col, out)$ is a good triple, $col_i=out_j$ by the third condition in the definition of good triples. Then $c''(v)=col_i$, a contradiction. 
			
			If the unsatisfied restrictions are imposed by $out'$, then $v \in N_G(V_i')\setminus N_{G_{\Phi'}}(V_i')$ for some $i$, $out_i'\neq 0$ and $c''(v)\neq out_i'$. If $v \in V''_j$ for some $j$, then $col''_j=col_j=out_i=out'_i$ by the third condition of a good triple, and we obtain a contradiction like in the previous case. Thus, $v \notin V(G_{\Phi''})$, so $v \in N_G(V''_j)\setminus N_{G_{\Phi''}}(V''_j)$ for some $j$, and $c''(v)=out''_j$. Since $N_G(V'_i)\setminus N_{G_{\Phi'}}(V'_i)=N_G(V_i)\setminus N_{G_{\Phi}}(V_i)$ and $N_G(V''_j)\setminus N_{G_{\Phi''}}(V''_j)=N_G(V_j)\setminus N_{G_{\Phi}}(V_j)$, and $(\Phi, col, out)$ is a good triple, by the fourth condition of a good triple, $out_i=out_j$. Hence, $out'_i=out''_j=c''(v)$, a contradiction. We finally obtain that $c \in \mathcal{C}(\Phi', col', out')\cap \mathcal{C}(\Phi'', col'', out'')$.
			
			(ii) Straightforwardly follows from Claim~\ref{claim5} and (i).
			
			(iii) By definition of $c$, all vertices in $V(G_{\Phi''})$ that are happy (and counted in $OPT(\Phi'', col'', out'')$) with respect to $c''$ in $G$, are happy with respect to $c$ as well. Thus, $\mathcal{H}(G,c)\supseteq \mathcal{H}(G, c'') \cap \mathcal{V}(\Phi'', out'')$. It is left to show that $c(v)=c'(v)$ for each $v \in N_G[\mathcal{H}(G,c')\cap \mathcal{V}(\Phi', col')]$. That is, $c$ preserves colors of all happy vertices in $\mathcal{V}(\Phi', col')$ and all their neighbours in $G$, with respect to $c'$.
			
			Suppose that it is not true, and there is a vertex $v \in N_G[\mathcal{H}(G,c')\cap \mathcal{V}(\Phi', col')]$ such that $c(v)\neq c'(v)$. Then $v \in V(G_{\Phi''})\cup N_G(\mathcal{V}(\Phi'', out''))$ and $c'(v)\neq c''(v)$. Using the fact that $(\Phi, col, out)$ is a good triple, we can obtain a contradiction. We skip this case analysis, as it is very similar to the case analysis shown above for proving that $c \in \mathcal{C}(\Phi', col', out')\cap \mathcal{C}(\Phi'', col'', out'')$.
		\end{claimproofO}
	
		Claim \ref{claim:mhv_cliquewidth_merge_colorings} shows that two colorings $c' \in \mathcal{C}(\Phi', col', out')$ and $c'' \in \mathcal{C}(\Phi'', col'', out'')$ that agree with $(\Phi, col, out)$ always can be merged in a single coloring $c \in \mathcal{C}(\Phi, col, out)$ of $G_\Phi$ preserving happy vertices in both $\mathcal{V}(\Phi',out')$ and $\mathcal{V}(\Phi'',out'')$, so Lemma \ref{lemma:mhv_cw_disj_main_equ} holds.
		With Lemma \ref{lemma:mhv_cw_disj_main_equ}, it is easy to compute the value of $OPT(\Phi, col, out)$ in $(\ell+1)^{2w} \cdot n^{\O(1)}$ time, as $|P_i(col_i)| \le (\ell+1)^2$.

		\item $\Phi = \rho_{i \to j} \Phi'$.
		\ifshort
\else	
		Take a coloring $c \in \mathcal{C}(\Phi, col, out)$.
		Note that necessarily $col_i=out_i=0$, as $V_i=\emptyset$.
		We want to find the values of $col'$ and $out'$, so that $c \in \mathcal{C}(\Phi', col', out')$ and $\mathcal{V}(\Phi, out)=\mathcal{V}(\Phi',out')$.
		As only labels $i$ and $j$ are touched by the topmost operator in $\Phi$, $col_k=col'_k$ and $out_k=out'_k$ for any $k$ not equal to $i$ or $j$.
		We assume that neither $V'_i$ nor $V'_j$ are empty, otherwise finding $col'$ and $out'$ is trivial.
		If $col_j=0$, i.e.\ $|c(V_j)|=|c(V'_i\cup V'_j)|\ge 2$, then necessarily at least one of $col_i$ and $col_j$ is equal to $0$ (so $|c(V'_i)|\ge 2$ or $|c(V'_j)| \ge 2$) or $col_i\neq col_j$ (so $|c(V'_i)\cup c(V'_j)|= 2$).
		If $col_j\neq 0$, then $col'_i=col'_j=col_j$ obviously.
		
		Now consider handling $out'_i$ and $out'_j$.
		If $out_j=0$ and $N_G(V_j)\neq N_{G_\Phi}(V_j)$, then the vertices with label $j$ in $G_\Phi$ are not counted in $OPT(\Phi,col,out)$, so we can skip counting them in $OPT(\Phi',col',out')$ by putting $out'_i=out'_j=0$.
		If $out_j\neq 0$, then $c(N_{G}(V_j)\setminus N_{G_\Phi}(V_j))=c(N_{G}(V'_i \cup V'_j)\setminus N_{G_{\Phi'}}(V'_i\cup V'_j))=out_j$, so $out'_i=out'_j=out_j$ necessarily.
		Hence, in either case $out'_i=out'_j=out_j$, so $out'=out$.
		
		Thus, we obtain
		\fi
		$$OPT(\Phi, col, out)=\max_{col'} OPT(\Phi', col', out),$$
		where the values of $col'_i$ and $col'_j$ are iterated over a few options (if $col_j\neq 0$, they are both equal to $col_j$, otherwise there are at most $({\ell+1})^2-\ell$ options as discussed above).
		For each $k$ not equal to $i$ or $j$, $col'_k=col_k$.
		
		\item $\Phi = \eta_{i,j} \Phi'$.
		\ifshort
		\else
		Again, take a coloring $c\in \mathcal{C}(\Phi,col,out)$ and consider finding appropriate $col'$ and $out'$ so that $c\in \mathcal{C}(\Phi,col',out')$ and $OPT(\Phi,col,out)=OPT(\Phi,col',out')$.
		
		Trivially, $col'=col$ as $V_i=V'_i$ for each $i \in [w]$.
		For each $k$ not equal to $i$ or $j$, it is enough to put $out'_k=out_k$, as $N_{G_{\Phi'}}(V'_k)=N_{G_\Phi}(V_k)$.
		
		Consider now the value of $out_i$.
		If $V_j$ is empty (hence, $col_j=0$), $G_\Phi$ equals $G_{\Phi'}$, so $out_i=out'_i$.
		If $out_i=0$, then we should put $out'_i=0$ so $\mathcal{V}_i(\Phi, out)=\mathcal{V}_i(\Phi', out')$.
		Suppose that $out_i\neq 0$ and $V_j$ is not empty.
		If $col_j=0$ or $col_j\neq out_i$, then all vertices in $V_i$ are unhappy with respect to $c$, as $V_j \sqcup (N_G(V_i)\setminus N_{G_\Phi}(V_i)) \subseteq N_{G}(V_i)$, but $|c(V_j \sqcup (N_G(V_i)\setminus N_{G_\Phi}(V_i)))|\ge 2$.
		Thus, we would like not to count the vertices in $V_i$ happy in $G_\Phi$, so we can put $out'_i=0$.
		In case when $col_j=out_i$, all vertices with label $i$ that are happy in $G_{\Phi'}$ with respect to $c$ are happy in $G_\Phi$ with respect to $c$ as well.
		Hence, one should put $out'_i=out_i$ in this case.
		The value of $out'_j$ is handled in the same way.
		
		We finally obtain that
		\fi $OPT(\Phi,col,out)=OPT(\Phi',col,out')$, where $out_k=out'_k$ for each $k$ not equal to $i$ or $j$, and
		$$
		\begin{matrix}
		out'_i=\left\{
		\begin{tabular}{@{}ll@{}}
			$0$, &\text{ if $out_i\neq 0$ and $col_j \neq out_i$},\\
			$out_i$,& \text{ otherwise;}
		\end{tabular}
		\right.&
		out'_j=\left\{
		\begin{tabular}{ll}
		$0$, &\text{ if $out_j\neq 0$ and $col_i \neq out_j$},\\
		$out_j$,& \text{ otherwise.}
		\end{tabular}
		\right.
		\end{matrix}$$
		This exhausts the list of possible cases.
	\end{enumerate}

	It is easy to see that the time required for the computation of a value $OPT(\Phi,col,out)$ requires polynomial time for each case, except for the case of the disjoint union operator.
	For this case, at most $(\ell+1)^{2w} \cdot n^{\O(1)}$ operations are required for the computation of a single cell of $OPT$.
	Since $OPT$ consists of at most $|\Psi| \cdot (\ell+1)^{2w}$ cells, computation of all values of $OPT$ takes at most $(\ell+1)^{4w} \cdot n^{\O(1)}$ running time.
	
	It is left to answer the initial problem question using the computed values of $OPT$.
	Note that each full coloring of $G$ extending $p$ is contained in $\mathcal{C}(\Psi,col,0)$ for some choice of $col$.
	Furthermore, $\mathcal{V}(\Psi,0)=V(G)$, so $OPT(\Psi,col,0)$ is the maximum number of happy vertices that can be obtained in $G$ with respect to colorings in $\mathcal{C}(\Psi,col,0)$.
	Thus, the maximum number of happy vertices that can be obtained in $G$ equals $\max_{col} OPT(\Psi,col,0)$, and this can be found in $\O((\ell+1)^{w})$ running time having all values of $OPT$ computed.
	This finishes the description of the algorithm.
\end{proofO}

\begin{corollary}
	\MHVfull~parameterized by $\operatorname{cw}+\ell$ admits an \FPT-algorithm.
\end{corollary}

In the rest of this section we show that \MHVfull~is polynomially solvable on the class of interval graphs, that is related to clique-width in the following sense.
Interval graphs have unbounded clique-width, moreover, unit interval graphs are minimal hereditary graph class of unbounded clique-width \cite{lozin2007clique}.
Since threshold graphs are a subclass of interval graphs, this result also covers the result of Choudhari and Reddy in \cite{Choudhari2018}, where they showed that \MHV~is polynomially solvable on the class of threshold graphs.
We also note that \MHE, in contrast to \MHV, is \NP-hard on the class of interval graphs, which is a corollary of Theorem \ref{thm:mhe_np_hard}.

\ifshort
\else
We start with the following convenient characterization of interval graphs.
\begin{theorem}[\cite{Gilmore1964}]\label{thm:interval_char}
	A graph is an interval graph if and only if its maximal cliques
	can be linearly ordered in such a way that for every vertex in the graph the maximal cliques to which it belongs occur consecutively in the linear order.
\end{theorem}

The sequence of the maximal cliques of an interval graph $G$ in the correct ordering from Theorem \ref{thm:interval_char} can be found in $\O(|V(G)|^2)$ time using the LBFS algorithm of Corneil, Olariu and Stewart \cite{Corneil2010}.
The following lemma is a folklore technical result, so it is given without a proof.

\begin{lemma}\label{lemma:interval_sequence}
	Let $G$ be an interval graph, $n=|V(G)|$, $m=|E(G)|$.
	There is a sequence $S_0, S_1, \ldots, S_{2n}$ of subsets of $V(G)$, such that
	\begin{itemize}
		\item $S_0=S_{2n}=\emptyset$;
		\item $uv \in E(G)$ if and only if $u, v \in S_i$ for some $i$;
		\item for each $i \in [2n-1]$, either $S_{i+1}=S_i \cup \{v\}$ for some $v \in V \setminus S_i$, or $S_{i+1} = S_i \setminus \{v\}$ for some $v \in S_i$;
		\item for each vertex $v \in V(G)$, $\{i: v \in S_i\}=[l_v,r_v-1]$ for some $0 < l_v < r_v \le 2n$.
	\end{itemize}
	Moreover, this sequence can be found in $\O(n^2)$ time.
\end{lemma}

We shall now prove a very useful property of this sequence.

\begin{lemmaO}\label{lemma:interval_clique_ordering}
	Let $S_1, S_2, \ldots, S_{2n}$ be the sequence from Lemma \ref{lemma:interval_sequence}.
	Let $i$ be an arbitrary integer in $[0, 2n]$.
	Let $G_i=G[S_0 \cup S_1 \cup S_2 \cup \ldots \cup S_i]$.
	There is an ordering $v_1, v_2, \ldots, v_{|S_i|}$ of the vertices in $S_i$ such that $N_{G_i}[v_1]\supseteq N_{G_i}[v_2] \supseteq \ldots \supseteq N_{G_i}[v_{|S_i|}]$.
	Moreover, this ordering can be found in $\O(n)$ time.
\end{lemmaO}
\begin{proofO}
	We claim that for each $S_i$ an appropriate order is the order $v_1, v_2, \ldots, v_{|S_i|}$ so that the values of $l_{v_i}$ go in the increasing order.
	As the values of $l_v$ are found in $\O(n)$ time, it is easy to find such ordering in $\O(n)$ time as well, using additional $\O(n)$ memory.
	
	It is easy to prove that this ordering is sufficient by induction on $i$.
	The base case $i=0$ is trivial since $S_i=\emptyset$.
	Let now $i>0$ be an integer and the claim hold for $i-1$.
	There are two possible cases: either $S_i=S_{i-1}\setminus \{v\}$ or $S_i=S_{i-1}\cup \{v\}$.
	In the former case, $G_{i-1}$ and $G_i$ are the same, and the ordering of the vertices $S_i$ is just a subsequence of the ordering of the vertices of $S_{i-1}$, so the claim holds true for $i$.
	In the latter case, $G_i$ differs from $G_{i-1}$ in the vertex $v$ and edges connecting each vertex in $S_{i-1}$ with $v$, so $N_{G_i}[v]=S_{i}$ and $N_{G_i}[u]=N_{G_{i-1}}[u]\cup \{v\}$ for each vertex $u \in S_{i-1}$.
	Since $S_{i}$ induces a clique in $G_i$, $S_i \subseteq N_{G_i}[u]$, equivalently, $N_{G_i}[v] \subseteq N_{G_i}[u]$ for each $u \in S_i$.
	As $v$ has the largest value of $l_v$ among all vertices in $S_i$, it stands the last in the ordering for $S_i$, and its neighbourhood is contained in the neighbourhood of each other vertex in $S_i$.
	The other vertices in the ordering of $S_i$ are no different from these of $S_{i-1}$, so the claim holds true for $i$ as well.
	The proof is finished.
\end{proofO}

We are now ready to present a polynomial time algorithm for the \MHVfull~problem on the class of interval graphs.
\fi
\begin{theoremO}\label{thm:mhv_interval}
	There is $\O(\ell n^2)$ running time algorithm for \MHVfull~on interval graphs.
\end{theoremO}
\begin{proofO}
	We present an algorithm solving \MHVfull~on the class of interval graphs.
	Let $(G, p, k)$ be an instance of \MHV~given to the algorithm, where $G$ is an interval graph, $n=|V(G)|$, $m=|E(G)|$.
	
	Firstly, the algorithm finds a sequence $S_0, S_1, S_2, \ldots, S_{2n}$ from Lemma \ref{lemma:interval_sequence} in $\O(n^2)$ time.
	Then it employs a dynamic programming over the sequence.
	Denote by $G_i$ the graph induced by the union of the first $i+1$ subsets in the sequence, i.e.\ $G_i=G[S_0 \cup S_1 \cup \ldots \cup S_i]$.
	Then, for each $i \in \{0\}\cup[2n]$, each $h \in \{0\}\cup [|S_i|]$, each $a \in \{-1\}\cup[\ell]$ and each $u \in \{-1\} \cup S_i$, the dynamic programming value $OPT(i, h, a, u)$ is formally defined as
	\begin{equation}\label{equ:interval_opt}
	OPT(i, h, a, u)=\max\left\{
	|\mathcal{H}(G_i, c)|\left|\,\,\begin{matrix}
	c:V(G_i) \to [\ell], \\
	\text{$c$ extends $p$ in $G_i$}, \\
	h=|\mathcal{H}(G_i, c) \cap S_i|, \\
	a=c\left(\argmax\limits_{v \in S_i}(r_v)\right),\text{ if }S_i\neq\emptyset;\text{ or }a=-1, \\
	
	u=\argmax\limits_{v \in S_i, c(v) \neq a}(r_v),\text{ if }|c(S_i)|>1;\text{ or }u=-1.
	\end{matrix}\right.
	\right\}.
	\end{equation}
	
	Also denote by $C(i,h,a,u)$ the set of all colorings of $G_i$ corresponding to the right part of equation \ref{equ:interval_opt} for $OPT(i, h, a, u)$, so $$OPT(i,h,a,u)=\max_{c \in C(i, h, a, u)}|\mathcal{H}(G_i, c)|.$$
	For each choice of $(i, h, a, u)$ such that there is no appropriate coloring $c$ for this choice, i.e.\ $C(i,h,a,u)=\emptyset$, we put $OPT(i, h, a, u)=-\infty$.
	Strictly speaking, $OPT(i, h, a, u)$ denotes the maximum number of vertices that can be happy simultaneously in $G_i$ with respect to colorings $c$ such that there are exactly $h$ happy vertices in $S_i$, the vertex $v \in S_i$ with the largest value of $r_v$ is colored with the color $a$.
	And $u$ denotes the vertex that is colored with a color different from $a$ with the largest value of $r_u$.
	
	The intuition behind this DP is the following.
	Since each $S_i$ induces a clique in $G$, $S_i$ can contain a happy vertex only if all its vertices are colored with the same color.
	Thus, we do not really need to store the colors of all vertices in $S_i$ as parameters of $OPT$, since happy vertices can be produced only when there is exactly one color in $S_i$.
	The value of $u$ allows to understand whether all vertices in $S_i$ are colored with the same color.
	Moreover, the value of $a$ clearly determines this color.
	Due to the interval structure of $G$, the values of $h$, $a$ and $u$ can be easily maintained while making transitions from $i$ to $i+1$.
	The following claim shows that the set of happy vertices inside $S_i$ is determined uniquely by the value of $h$.
	
	\begin{claim}\label{claim:happy_bag}
		Let $c \in C(i, h, a, u)$ be a coloring of $G_i$ extending $p$.
		Let $v_1, v_2, \ldots, v_{|S_i|}$ be an ordering of the vertices in $S_i$ from Lemma \ref{lemma:interval_clique_ordering}.
		Then $\mathcal{H}(G_i, c) \cap S_i=\{v_{|S_i|-h+1}, v_{|S_i|-h}, \ldots, v_{|S_i|}\}$.
	\end{claim}
	\begin{claimproof}
		By definition of $C(i, h, a, u)$, $|\mathcal{H}(G_i, c)\cap S_i|=h$.
		By Lemma \ref{lemma:interval_clique_ordering}, $N_{G_i}[v_j] \supseteq N_{G_i}[v_{j+1}]$ for each $j \in [|S_i|-1]$.
		Hence, if $v_j \in \mathcal{H}(G_i, c)$, then $v_{j+1}\in \mathcal{H}(G_i, c)$ also.
		Thus, $\mathcal{H}(G_i, c) \cap S_i$ consists of the last $h$ vertices from the ordering.
	\end{claimproof}
	
	Clearly, $OPT(0, 0, -1, -1)=0$ and every other value of $OPT(0, \cdot, \cdot, \cdot)$ equals $-\infty$, as $S_0=\emptyset$.
	Now the algorithm has all values of $OPT$ computed correctly for $i=0$.
	Then, the algorithm iterates over all values of $i$ from $0$ to $2n-1$.
	Having the value of $i$ fixed, our algorithm initializes all values of $OPT(i+1, \cdot, \cdot, \cdot)$ with $-\infty$.
	Then it iterates over each state of the dynamic programming $OPT(i,h,a,u)$ with $OPT(i,h,a,u)\neq -\infty$.
	Consider now a coloring $c \in C(i,h,a,u)$ of $G_i$.
	There are two possible options of how $G_{i+1}$ and $S_{i+1}$ differs from $G_i$ and $S_i$.
	If $S_{i+1}=S_i \setminus \{v\}$, then $G_{i+1}=G_i$, so $c$ is a full coloring of $G_{i+1}$ extending $p$.
	The following claim formalizes dynamic programming transitions that should be made in this case.
	
	\begin{claim}\label{claim:transitions_remove}
		Let $c \in C(i, h, a, u)$.
		Suppose $S_{i+1}=S_i \setminus \{v\}$ for some $v \in S_i$.
		Then $c\in C(i+1,h',a',u')$, where
		\begin{equation}\label{equ:transitions_remove_happy}
		h'=\left\{
		\begin{tabular}{ll}
		$h-1$, & if $v \in \mathcal{H}(G_i, c)$,\\
		$h$, & otherwise.
		\end{tabular}
		\right.
		\end{equation}
		and
		\begin{equation}
		a'=\left\{
		\begin{tabular}{ll}
		$-1$, & if $S_{i+1}=\emptyset$,\\
		$a$, & otherwise.
		\end{tabular}
		\right.
		\end{equation}
		and
		\begin{equation}
		u'=\left\{
		\begin{tabular}{ll}
		$-1$, & if $u=v$,\\
		$u$, & otherwise.
		\end{tabular}
		\right.
		\end{equation}
		and
		\begin{equation}
		\mathcal{H}(G_{i+1},c)=\mathcal{H}(G_i, c).
		\end{equation}
	\end{claim}
	\begin{claimproof}
		$\mathcal{H}(G_{i+1},c)=\mathcal{H}(G_i,c)$ is true since $G_i=G_{i+1}$.
		Recall that $h'=|\mathcal{H}(G_{i+1},c) \cap S_{i+1}|=|\mathcal{H}(G_i, c) \cap (S_{i}\setminus \{v\})|$.
		If $v \in \mathcal{H}(G_i,c)$, then $h'=h-1$, otherwise $h'=h$.
		
		Recall that $a$ denotes the color of the vertex $w \in S_i$ with the largest value of $r_w$, and $a'$ denotes such a color in $S_{i+1}=S_i \setminus \{v\}$.
		Since $v$ is the vertex with the smallest value of $r_v$ in $S_i$ ($r_v=i+1$), $a'$ should change only if $S_i=\{v\}$.
		The same can be said about $u$ and $u'$.
	\end{claimproof}
	
	Basically, Claim \ref{claim:transitions_remove} states that $C(i,h,a,u)=C(i+1,h',a',u')$, where $h'$, $a'$ and $u'$ depend only on the values of $i,h,a$ and $u$.
	Note that in order to compute the value of $h'$, our algorithm firstly finds $\mathcal{H}(G_i,c)\cap S_i$ as stated in Claim \ref{claim:happy_bag}.
	To do this, the algorithm needs the ordering of the vertices of $S_i$ from Lemma \ref{lemma:interval_clique_ordering}.
	This ordering is found once for a fixed value of $i$ in $\O(n)$ time.
	When the ordering for a fixed $i$ is known, then, by Claim \ref{claim:happy_bag}, the condition $v \in \mathcal{H}(G_i, c)$ (equivalently, $v \in \mathcal{H}(G_i, c) \cap S_i$) from equation \ref{equ:transitions_remove_happy} can be checked in $\O(1)$ time using the value of $h$.
	Thus, for each fixed values of $i,h,a$ and $u$, the algorithm computes $h',a'$ and $u'$ and updates current value of $OPT(i+1,h',a',u')$ (initially all values of $OPT(i+1,\cdot,\cdot,\cdot)$ are equal to $-\infty$) with the value of $OPT(i,h,a,u)$: $$OPT(i+1, h',a',u'):=\max\{OPT(i+1,h',a',u'),OPT(i,h,a,u)\}.$$
	Since each full coloring of $G_{i+1}$ is a full coloring of $G_i$ and all values $OPT(i,\cdot,\cdot,\cdot)$ are computed correctly, all values of $OPT(i+1,\cdot,\cdot,\cdot)$ are computed correctly as well.
	Note that if $h>0$ and $u \neq -1$, then necessarily $OPT(i,h,a,u)=-\infty$, as there may be no happy vertices inside $S_i$, if $S_i$ contains two vertices colored with distinct colors.
	The algorithm does not iterate over such values of $OPT(i,h,a,u)$, so for a fixed value of $i$ it iterates over at most 
	\begin{equation*}
	\label{eq:dynamic-size}
	\tag{**}
|S_i| \cdot \ell \cdot 1 + 1 \cdot \ell \cdot |S_i|=\O(\ell n)
	\end{equation*} values.
	For each fixed triple of values of $h,a$ and $u$ the transitions are done in $\O(1)$ time, so it takes $\O(\ell n)$ time in total to compute all values of $OPT(i+1,\cdot,\cdot,\cdot)$ in the case $S_{i+1}=S_i \setminus \{v\}$.
	
	We now describe transitions for the second case, when $S_{i+1}=S_i \cup \{v\}$, so $G_{i+1}$ differs from $G_i$ by a vertex $v$ and all edges between $v$ and $S_i$.
	Now, for a coloring $c \in C(i, h, a, u)$ we consider each extension of $c$ onto $v$ agreeing with $p$.
	That is, if $v$ is a vertex precolored by $p$, we consider only one extension of $c$.
	Otherwise, we consider all $\ell$ colors of $v$ in an extension of $c$.
	When the color of $v$ in an extension of $c$, say $c'$, is specified, then $c' \in C(i+1,h',a',u')$ for some values of $h'$, $a'$ and $u'$.
	The following claim shows how to determine these values.
	
	\begin{claim}\label{claim:transitions_add}
		Let $c \in C(i, h, a, u)$. Suppose $S_{i+1}=S_i \cup \{v\}$ for some $v \in V \setminus S_i$ and denote $w=\argmax\limits_{w \in S_i} r_w$, if $S_i\neq \emptyset$.
		Let $c'$ be an extension of $c$ onto $V(G_{i+1})$ such that $c'(v)=b$.
		Then $c' \in C(i+1, h', a', u')$, where
		\begin{equation}
		h'=\left\{
		\begin{tabular}{ll}
		$h+1$, & if $a=-1$ or ($a=b$ and $u=-1$),\\
		$0$, & otherwise.
		\end{tabular}
		\right.
		\end{equation}
		and
		\begin{equation}
		a'=\left\{
		\begin{tabular}{ll}
		$a$, & if $a \neq -1$ and $r_v < r_w$, \\
		$b$, & otherwise.
		\end{tabular}
		\right.
		\end{equation}
		and
		\begin{equation}
		u'=\left\{
		\begin{tabular}{ll}
		$-1$, & if $u=-1$ and ($a=-1$ or $a=b$),\\
		$u$, & if $a=b$ or ($u \neq -1$ and $r_v < r_u$),\\
		$w$, & if $a \neq b$ and $r_v > r_w$,\\
		$v$, & otherwise.
		\end{tabular}
		\right.
		\end{equation}
		and
		\begin{equation}
		|\mathcal{H}(G_{i+1}, c')|=|\mathcal{H}(G_i, c)|+h'-h.
		\end{equation}
		
	\end{claim}
	\begin{claimproof}
		Consider proving the equation for $h'$.
		Recall that $N_{G_{i+1}}(v)=S_i$ and $h'=|\mathcal{H}(G_{i+1},c') \cap S_{i+1}|=|\mathcal{H}(G_{i+1},c') \cap (S_{i}\cup\{v\})|$.
		Note that if $v$ is happy with respect to $c'$, then all vertices in $S_i$ are colored with the color $b$ or $S_i=\emptyset$.
		Clearly, $S_i=\emptyset$ is equivalent to $a=-1$.
		Note that $u=-1$ is equivalent $|c(S_i)|\le 1$.
		Thus, $u=-1$ and $a=b$ is equivalent to that $S_i$ is not empty and all vertices in $S_i$ are colored with the color $b$ by $c$.
		Thus, if $v$ is happy with respect to $c'$ in $G_{i+1}$, then $\mathcal{H}(G_{i+1},c')=\mathcal{H}(G_i,c)\cup \{v\}$, so $h'=h+1$.
		If, otherwise, $v \notin \mathcal{H}(G_{i+1},c')$, then no vertex in $S_i$ (hence no vertex in $S_{i+1}$), is happy with respect to $c'$ in $G_{i+1}$, so $h'=0$.
		The equality $|\mathcal{H}(G_{i+1}, c')|=|\mathcal{H}(G_i, c)|+h'-h$ follows from the discussion above and the fact that $\mathcal{H}(G_{i+1}, c')\setminus S_{i+1}=\mathcal{H}(G_{i}, c)\setminus S_{i}$.
		
		To prove the equation for $a'$, recall that $$a'=c'\left(\argmax_{z\in S_{i+1}}(r_z)\right)=c'\left(\argmax_{z\in S_{i}\cup \{v\}}(r_z)\right)=c'\left(\argmax_{z\in \{v,w\}}(r_z)\right),$$
		if $S_i\neq \emptyset$.
		Hence, $a'=c'(w)=c(w)=a$ if $r_w > r_v$, and $a'=c'(v)=b$ otherwise.
		If $S_i=\emptyset$, which is equivalent to $a=-1$, then $S_{i+1}=\{v\}$, so obviously $a'=c'(v)=b$.
		
		It is left to prove the equation for $u'$.
		Recall that $u$ stands for the vertex in $S_{i}$ with the largest value of $r_{u}$ among all vertices that are colored with a color different from $a$.
		$u'$ denotes the same vertex, but for $S_{i+1}$ and $a'$.
		If $u=-1$, then either $S_i$ is empty (equivalently, $a=-1$) or all vertices in $S_i$ are colored with the same color $a$.
		If $a=-1$ or $a=b$, then $S_{i+1}$ is a set of vertices colored with the same color.
		Hence, in this case $u'=-1$.
		For any other case, $u' \neq -1$.
		The other parts of the equation for $u'$ are handled technically in a similar way.
	\end{claimproof}
	
	Similarly to the previous case, algorithm iterates over all values of $h, a$ and $u$ (except for values with $OPT(i,h,a,u)=-\infty$), computes the values of $h', a', u'$ according to Claim \ref{claim:transitions_add} and performs an update $$OPT(i+1,h',a',u'):=\max\{OPT(i+1,h',a',u'),OPT(i,h,a,u)+h'-h\}.$$
	Note that the value of $w$ in Claim \ref{claim:transitions_add} depends only on $i$ and is computed once in $\O(n)$ time.
	Also in this case algorithm also has to iterate over all $\ell$ values of $c'(v)$ (except for if $p(v)$ is specified, then it is just a single fixed value).
	Iterating over the values of $h,a$ and $u$ takes $\O(\ell n)$ time by~\ref{eq:dynamic-size} and iterating over the value of $c'(v)$ takes $\O(\ell)$ time and the transitions are done in $\O(1)$ time, so the computation of all values of $OPT(i+1,\cdot,\cdot,\cdot)$ takes $\O(\ell^2 n)$ time in total.
	
	This already gives us an overall $\O(\ell^2 n^2)$ running time bound for the algorithm.
	We shall now optimize the algorithm to obtain the upper bound of $\O(\ell n^2)$ for its running time.
	Note that for the values of $i$ with $S_i=\emptyset$ the running time bound of $\O(1)$ is achieved by the algorithm, as the only value of $OPT(i,\cdot,\cdot,\cdot)$ not equal to $-\infty$ is $OPT(i,0,-1,-1)$.
	We now assume that $S_i\neq \emptyset$, hence $a \neq -1$.
	Consider the case when $r_v > r_w$ and the equations in Claim \ref{claim:transitions_add}.
	Note that the values of $h', a'$ and $u'$ do not depend on the value of $b$ itself, but only on whether $a \neq b$ or not.
	Thus, it is not necessary for the algorithm to iterate over all $\ell$ possible values of $b=c'(v)$, as all values of $c'(v)$ different from $a$ are handled in the same way with the same dynamic programming transitions.
	So it is enough for the algorithm to consider only two cases $c'(v)=a$ and $c'(v)\neq a$ (if $v$ is a precolored vertex, there is still the only case $c'(v)=p(v)$).
	Hence, for the values of $i$ for which $r_v < r_w$, this change in the algorithm achieves the desired $\O(\ell n)$ running time bound.
	
	Consider now the other case, $r_v > r_w$.
	In this case, the value of $a'$ equals to the color of $v$ in $c'$, but does not depend on $a$.
	It only matters whether $a\neq b$ or not for correct computation of $h'$ and $u'$.
	Thus, in this case algorithm iterates over possible values of $h$, $u$ and $c'(v)$, but not fixing the value of $a'$.
	To handle the case $c'(v)=a$, it is enough to find the values of $h'$ and $u'$ accordingly to Claim \ref{claim:transitions_add} and make the transition $$OPT(i+1,h',c'(v),u'):=\max\{OPT(i+1,h',c'(v),u'),OPT(i,h,c'(v),u)+h'-h\}.$$
	To handle the other case, $c'(v)\neq a$, essentially a transition of kind 
	$$OPT(i+1,h',c'(v),u'):=\max\{OPT(i+1,h',c'(v),u'),\max_{a\neq c'(v)}OPT(i,h,a,u)+h'-h\}$$
	is enough to be made.
	Thus, it is enough to compute $\max_{a\neq c'(v)}OPT(i,h,a,u)$ faster for any given value of $c'(v)$.
	To achieve that, for a fixed triple of values $i,h$ and $u$ compute a sequence $p_0, p_1, \ldots, p_\ell$, where $p_0=-\infty$ and $p_{j}=\max\{p_{j-1}, OPT(i,h,j,u)\}$ for each $j \in [\ell]$, in $\O(\ell)$ time straightforwardly. $p_j$ is essentially the maximum value among the values of $OPT(i,h,\cdot,u)$ taken for the first $j$ colors.
	Analogously compute a sequence $s_{\ell+1},s_{\ell},\ldots,s_1$, where $s_{\ell+1}=-\infty$ and $s_j=\max\{s_{j+1},OPT(i,h,j,u)\}$ for each $j \in[\ell]$, in $\O(\ell)$ running time.
	In other words, $s_j$ is the maximum value of $OPT(i,h,\cdot,u)$ among the colors from $j$ to $\ell$.
	When these two sequence are computed, then, clearly, $\max_{a\neq c'(v)}OPT(i,h,a,u)$ can be found as $\max\{p_{c'(v)-1},s_{c'(v)+1}\}$.
	Hence, the transition above can be done in $\O(1)$ time.
	For a fixed value of $i$, the sequences $p$ and $s$ are computed in $\O(\ell)$ time once for each pair of values of $h$ and $u$, then, for each triple $h,u$ and $c'(v)$, the transition is done in $\O(1)$ time.
	Thus, the algorithm requires $\O(\ell n)$ running time for these steps.

	This finishes the description of the computation of $OPT$.
	For each fixed value of $i$ the computation of $OPT(i+1,\cdot,\cdot,\cdot)$ takes $\O(\ell n)$ time, so it takes $\O(\ell n^2)$ time in total to compute all values of $OPT$.
	
	Clearly, the maximum number of happy vertices that can be obtained in $(G,p)$ equals $OPT(2n,0,-1,-1)$, as $G_{2n}=G$ and $C(2n,0,-1,-1)$ equals the set of all colorings of $G$ extending $p$.
	Hence, the algorithm finally checks that $OPT(2n,0,-1,-1)$ is at least $k$ to determine whether $(G,p,k)$ is a yes-instance of \MHV.
	This finishes the proof.	
\end{proofO}
	\bibliographystyle{splncs04}
	\bibliography{ref}
\end{document}